\documentclass[12pt]{article}
\usepackage{psfrag,epsfig}
\usepackage{a4,isolatin1,amsmath}
\newcommand{\gtsim}
{\mathrel{\vbox{\hbox{$>$}\vskip-1.8ex\hbox{$\sim$}\vskip-.5ex}}}
\begin{document}
\begin{center}
{\bf \Large Transition from 
%photon 
antibunching to bunching for two dipole-interacting atoms
% Photon Bunching and AntiBunching for Two Dipole-Interacting
%  Atoms 
\normalsize}

\vspace*{0.5cm}

\noindent {\bf Almut Beige and Gerhard C. Hegerfeldt}\\
Institut für Theoretische Physik\\
Universität Göttingen\\
Bunsenstr. 9\\
37073 Göttingen, Germany
\end{center}

{\abstract
It is known that  there is a
transition from photon antibunching to bunching in the resonance
fluorescence of a  driven system
of two two-level atoms with dipole-dipole interaction when the atomic distance
decreases and the other parameters are kept fixed. 
We give a simple explanation for the underlying mechanism
which in principle can also be applied to other systems.\\[0.2cm]
PACS numbers 42.50.Ar, 42.50Fx}
\section{Introduction}
Bunching means that photons emitted by a driven system in 
steady state have a
tendency to arrive in pairs or larger groups at a detector rather than
uniformly distributed in time. More precisely, right after a photon emission
the probability density for emitting another photon is larger than for
a uniform distribution of corresponding emission rate. For antibunching
this probability density is smaller than for a uniform distribution,
and it means that the photons repel each other.

For a driven two-level system it is known that one has antibunching 
\cite{CarWa,KiDaMa,DiWal87}. This is intuitively very simple to understand since
after an emission the system is in its ground state and it requires
some time to acquire enough  population of the excited state for a
next emission. For two
independent, noninteracting, two-level atoms one also has
antibunching, although not quite so pronounced as for a single
atom. This was investigated experimentally in Ref.
\cite{Itano2}. 

For a system of two two-level atoms with dipole-dipole interaction it
is known from  studies of the master or optical Bloch equations for
the system that the properties of the resonance fluorescence may
change considerably when compared to that for a single system 
\cite{aga}-\cite{Berman}. In particular there is a
transition from antibunching to bunching when the atomic
distance becomes small and the other parameters are kept fixed 
(see e.g. Ref. \cite{FiTaKi83}). The state space of the system
is four-dimensional  and the density matrices have 16 components so
that the corresponding Bloch equations 
require diagonalization of a 16 $\times$ 16 matrix. This makes an 
intuitive understanding of this transition from antibunching to
bunching not obvious.

It is the aim of this paper to elucidate the underlying reason and
give a simple explanation for the appearance of bunching for small
distances in a driven system of two two-level atoms  with
dipole-dipole interaction. We will trace the phenomenon to two
causes. One is the form of the steady-state density matrix of the
system.  The other is the density
matrix of the system right after an emission, called the reset matrix 
\cite{He}, which depends only on the state prior to the emission.
In it the steady-state ground-state population has
disappeared since it does
not contribute to an emission, while the steady-state 
populations of the higher
levels have been transferred to  levels one step lower, in proportion to
their respective decay constants, and then normalized to 1. 
The new populations of the intermediate states then determine
the probability density for the next emission.

The transition from antibunching to bunching 
can be directly read off from the steady-state density
matrix of the system without a lengthy calculation.  For small atomic
distances and for  the Rabi
frequency  of the order of the Einstein coefficient $A$ or a few times larger
the ground-state population is  very large  while the excited
states have small but similar populations,  
and thus there  is only a small steady-state emission
rate. The 
large ground state population does not contribute to a photon emission and   
therefore disappears after an emission, while the populations of the
intermediate  states move to the ground state and that of the highest
state to the intermediate ones, in proportion to their
respective decay constants  and with ensuing normalization.  
Hence after a photon emission the ground and excited
states have suddenly acquired populations of the same order of
magnitude and thus a higher emission probability density than before. This
means that the photons tend to come in pairs and that there is
bunching. For weaker driving and  small
distances the same mechanism holds. In this case
the  population of the ground state is even larger and that of the
highest excited state  smaller than for the other excited states. But
its population  is still 
large enough so that after an emission, when the populations have moved
one step down, the new population of the intermediate states is
larger than before so that 
there is an increased emission probability and thus bunching. 
These considerations can be applied to quite general systems.

Here we consider  a system of two two-level
atoms at a fixed distance $r$, interacting with the quantized radiation
field and a classical laser field. Through photon exchange the 
radiation field mediates the
$r$ dependent dipole-dipole interaction of the atoms. The dipole
and rotating-wave approximation is used throughout. Retardation effects are
included.

In Section 2 we briefly review the  photon-counting correlation
function $g(\tau)$.  If $g(0) > 1$ one has bunching, if 
$g(0) < 1$ one has antibunching. 
In Section 3 we apply this to two two-level
atoms and give an explicit  expression for $g(0)$ as a function of the
atomic distance and the driving field.
Bunching is explicitly seen 
for atomic distances about a quarter of a
wavelength or less.

In Section 4 we discuss the results and the simple mechanism
responsible for bunching for small atomic distances. 
It is also made clear why an analogous argument yields antibunching 
for two independent non-interacting atoms and for
atoms  sufficiently far away.

For simplicity we consider, in the main part of the paper, coinciding
 atomic dipole moments and the same laser phase for each atom.
In the Appendix the general case is considered, the corresponding
 Hamiltonian spelled out and the reset
 matrix given. 
We also outline the connection with  the quantum jump approach
\cite{HeWi,Wi,He,HeSo}  which is equivalent to the Monte Carlo
wave-function approach \cite{MC} and to quantum trajectories
\cite{QT}. For a recent review of this approach see Ref. \cite{PleKni}. 
For a system of two two-level atoms with dipole-dipole interaction we 
carry this approach over in a form convenient for simulations . 
%Kap. 1 
\section{Photon-counting correlation functions}
%Kap. 1
We briefly review some well-known facts. 
As pointed out in Ref. \cite{MaWo} 
there is a minor difference between  correlation functions which are 
based on the electric field operator and those which are based on the
photon number operator. Here we employ the latter type correlation
functions of second order. 
For simplicity we consider broad-band photon detections
over all space. It is useful to distinguish clearly between  correlation
functions for ensembles and  for a
single  trajectory. The latter involves a time rather than an
ensemble average. 

{\em Ensemble}. 
Consider an ensemble of laser driven atomic systems in the state
$\rho$ at $t = 0$ and denote by $G(t_2, t_1; \rho) dt_1 dt_2$ the
relative number of systems for which in addition to a photon in $(t_1,
t_1 + d t_1)$ also a photon in $(t_2, t_2 + dt_2)$ is detected. If,
for a particular trajectory, we denote the number of photons detected
in $(t, t + \Delta t)$ by $N^{{\rm traj}}(t, t + \Delta t)$ -- for
small $\Delta t$ this
number is either 0 or 1  -- then 
\begin{equation}\label{34a}
G(t_2, t_1; \rho)dt_1 dt_2 = \langle N^{{\rm traj}}(t_1, t_1  + dt_1)
N^{{\rm traj}}(t_2, t_2 + d t_2)\rangle_{{\rm ens}}~.
\end{equation}
 Let us consider 
the sub-ensemble of systems which had an  emission at $t_1$ and let us
denote its normalized  density matrix  right after the emission
 by $\hat{{\cal R}}(\rho (t_1))$.
This we call the normalized reset
matrix and it will be given explicitly for  a two-atom system in the next
section. We denote by $I(t;\rho)$ the probability density 
for the emission of a photon at time $t$ (not necessarily the first photon
after $t =0$) for initial density matrix $\rho$. With this one has
\begin{equation}\label{34b}
G(t_2, t_1; \rho)dt_1 dt_2 = I(t_1 ; \rho) dt_1~I(t_2 - t_1 ; \hat{{\cal
    R}}(\rho (t_1))) dt_2~.
\end{equation}
Letting $t_1 \rightarrow \infty$ and keeping $\tau = t_2 - t_1$
fixed the first factor on the r.h.s. goes to $I_{{\rm s s}}$ and
$\rho(t_1)$ to $\rho^{{\rm s s}}$, the steady-state emission rate and
density matrix, respectively. Hence
\begin{equation}\label{34c}
G(\tau) \equiv \lim_{t \rightarrow \infty} G(t + \tau, t ; \rho) =
I_{\rm s s} I(\tau ; \hat{\cal R}(\rho^{{\rm s s}}))~.
\end{equation}
The  photon correlation function $g(\tau)$ is now defined as
\begin{equation}\label{34d}
g (\tau) \equiv~\frac{G(\tau)}{I_{\rm s s} I_{\rm s s}}~=~\frac{I (\tau ;
  \hat{{\cal R}}(\rho^{\rm s s}))}{I_{\rm s s}}~.
\end{equation}
It compares, for the steady state, the probability density for
emission of a photon at a time interval $\tau$ after a preceding
emission with that of a uniform distribution of emission rate  $I_{s s}$. 

{\em Single trajectory}.
We now consider a single system with its trajectory of photon
emissions and define $N^{{\rm traj}}$ as before. 
At instances $t_m' = m \Delta t'$ until time
$T = M \Delta t'$ one measures  whether or not a photon has been emitted in
$(t_m' , t'_{m+1})$. Then the relative frequency  of cases in
which both in $(t'_m , t'_m + \Delta t_1)$ and $(t'_m + \tau, t'_m
+ \tau + \Delta t_2)$ a photon has been found is given in the
limit $\Delta t' \rightarrow 0$ and $T \rightarrow \infty$, using 
$1/M=\Delta t'/T$, by 
\begin{eqnarray}
G^{{\rm traj}}(\tau) \Delta t_1 \Delta t_2 
&=& \lim_{T \to \infty}\lim_{\Delta t' \to 0} \, {\Delta t' \over T} \, 
\sum_m N^{\rm traj}(t_m',t_m'+\Delta t_1) N^{\rm traj}(t_m'+\tau
 ,t_m'+
\tau +\Delta t_2) \nonumber \\
&=& \lim_{T \to \infty} \, {1 \over T} \int_0^T {\rm d}t' \, 
N^{\rm traj}(t',t'+\Delta t_1) N^{\rm traj}(t'+\tau,t'+\tau +\Delta 
t_2)~. 
\end{eqnarray} 
By ergodicity this should be the same for each trajectory, and
therefore one can take the ensemble average of the r.h.s. without
changing anything. Using Eqs. (\ref{34a}) and (\ref{34c}) one then
obtains
\begin{equation}\label{34f}
G^{{\rm traj}}(\tau) = G (\tau)
\end{equation}
so that both correlation functions coincide and similarly $g^{{\rm
    traj}}(\tau) = g(\tau)$. 

We also point out the well-known fact that 
if one observes  photons with a detector of efficiency $\eta$ less than 1
then  in Eq. (\ref{34d})  both numerator and denominator are
multiplied by $\eta$ and hence $g(\tau)$ is not affected by the
detector efficiency.

One has bunching if the relative number of cases, in which shortly
after emission of a photon a further photon is emitted, exceeds those
for a uniform distribution of frequency $I_{\rm s s}$. Thus bunching means
$g(0) > 1$. Similarly one has antibunching if this number is less
than for a uniform distribution, i.e. if $g(0) < 1$.
%Kap. 2
\section{\bf Bunching for two atoms}
%Kap. 2
We now turn to two two-level atoms with dipole-dipole interaction,
driven by a laser tuned to the atomic transition frequency $\omega_{0}$. 
The corresponding Hamiltonian is given in the Appendix. For
simplicity we consider coinciding atomic dipole moments forming an
angle $\vartheta$ with the line connecting the atoms 
 and laser radiation normal to this line
so that the laser is in phase for both atoms.
The Rabi frequency of the laser, denoted by $\Omega$, is then the same
for both atoms.
One can take $\Omega$ to be real and positive. The general case is
indicated in the Appendix.

It is convenient to use the  Dicke states \cite{Dicke}
$| g \rangle = |1\rangle |1\rangle $, $
| e \rangle  =  | 2 \rangle |2 \rangle$, and
$| s \rangle$ and
$| a \rangle$  the symmetric and antisymmetric combinations of $| 1
\rangle | 2 \rangle$ and $| 2 \rangle | 1 \rangle$.
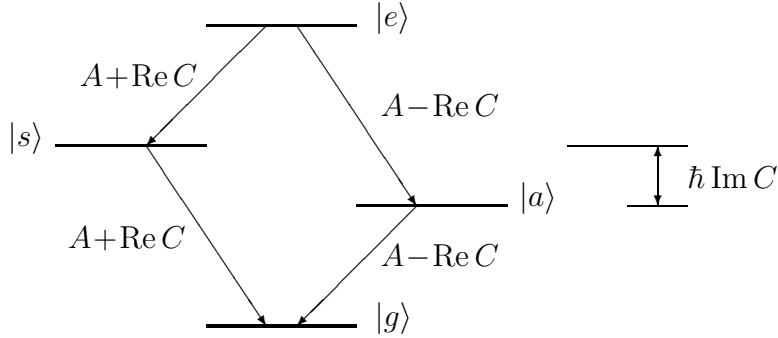
\begin{figure}[h] 
\hspace*{2cm}\unitlength0.8cm
\begin{picture}(14.5,6.5) 
\thicklines
\put(4,1){\line(1,0){2.5}}
\put(1.5,4){\line(1,0){2.5}}
\put(4,6){\line(1,0){2.5}}
\put(6.5,3){\line(1,0){2.5}}
\thinlines
\put(10,4){\line(1,0){2}}
\put(11,3){\line(1,0){1}}
\put(5,6){\vector(-1,-1){2}}
\put(5.5,6){\vector(2,-3){2}}
\put(3,4){\vector(2,-3){2}}
\put(7.5,3){\vector(-1,-1){2}}
\put(11.5,4){\vector(0,-1){1}}
\put(11.5,3){\vector(0,1){1}}
\put(0.7,4){$|s\rangle$}
\put(6.8,6){$|e\rangle$}
\put(9.2,3){$|a\rangle$}
\put(6.8,1){$|g\rangle$}
\put(1.9,5){$A \! + \! {\rm Re} \, C$}
\put(1.7,2.3){$A \! + \! {\rm Re} \, C$}
\put(6.9,4.5){$A \! - \! {\rm Re} \, C$}
\put(6.9,2){$A \! - \! {\rm Re} \, C$}
\put(12,3.3){$\hbar \, {\rm Im} \, C$}
\end{picture} \\[-1cm]
\caption{Dicke states and decay rates} 
\label{bild14}
\end{figure}
These states play the role of dressed states for the atoms
(cf. e.g. Ref. \cite{Brewer2}),
with decay constants  $A\pm{\rm Re}\,C$ (see Fig.\ref{bild14})
where $C$ is an  $r$ dependent  complex coupling constant. It is given
for the general case in Eq. (\ref{210}) of the Appendix.
From Fig. \ref{bild12b} it
is seen that $C \rightarrow 0$ for $r \rightarrow \infty$,
$|$Im$|\,C\rightarrow \infty$ for $r \rightarrow 0$, while Re$\, C$
changes little with $r$. Retardation effects are included in the
sense that  $C$ goes to its value for a static dipole-dipole 
interaction when  $c \rightarrow \infty$ \cite{VaAga}.
\begin{figure}[h]
\psfrag{w=0}{\hspace*{-0.3cm} {\tiny $\vartheta=0$}}
\psfrag{w=P/8}{\hspace*{-0.3cm} {\tiny $\vartheta=\pi/8$}}
\psfrag{w=P/4}{\hspace*{-0.3cm} {\tiny $\vartheta=\pi/4$}}
\psfrag{w=P/2}{\hspace*{-0.3cm} {\tiny $\vartheta=\pi/2$}}
\psfrag{Re C/A}{${\rm Re} \, C~[A]$}
\psfrag{Im C/A}{${\rm Im} \, C~[A]$}
\psfrag{a}{$k_0r$}
\hspace*{-0.9cm} \epsfig{file=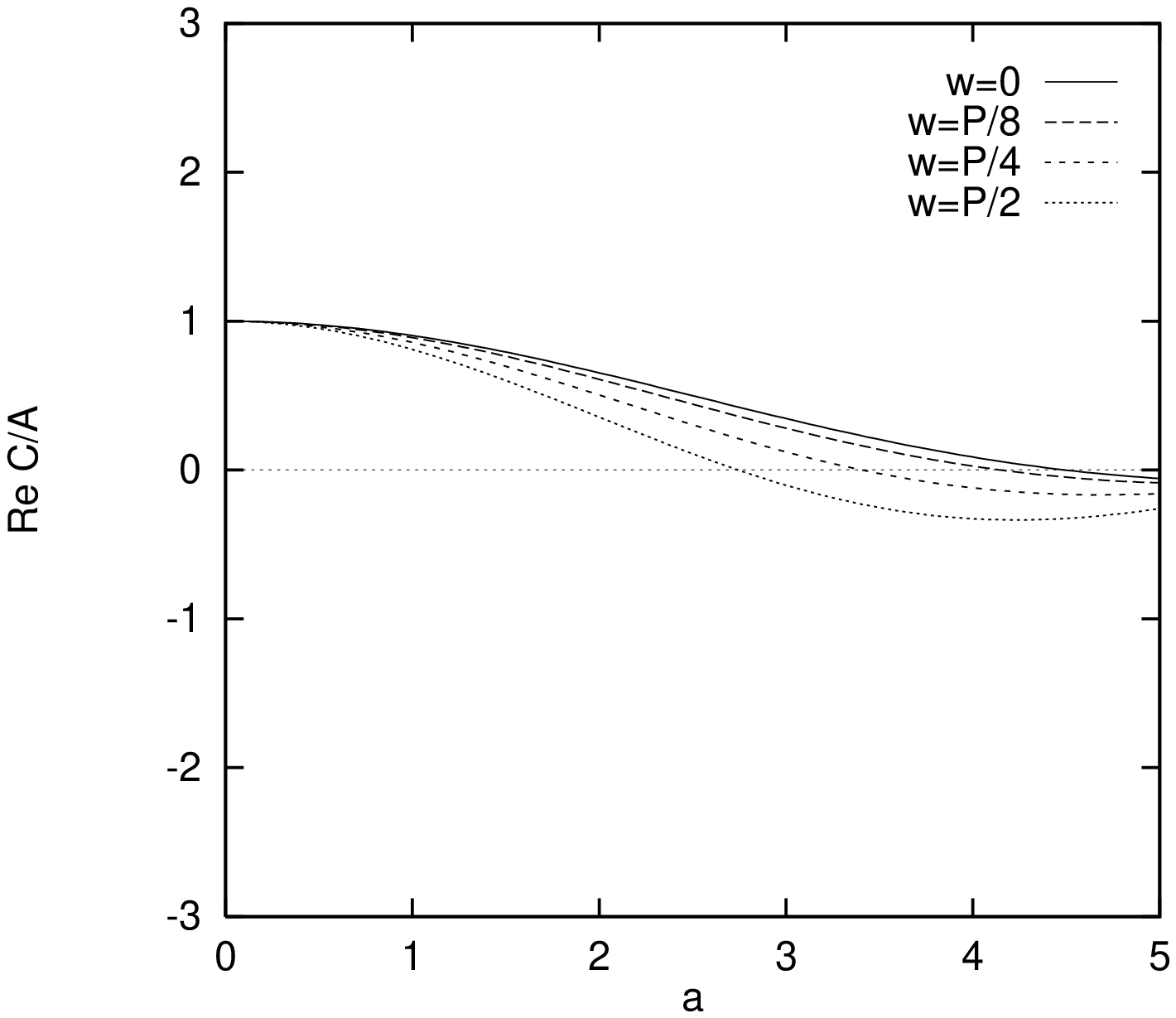, width=8.5cm} ~~~
\epsfig{file=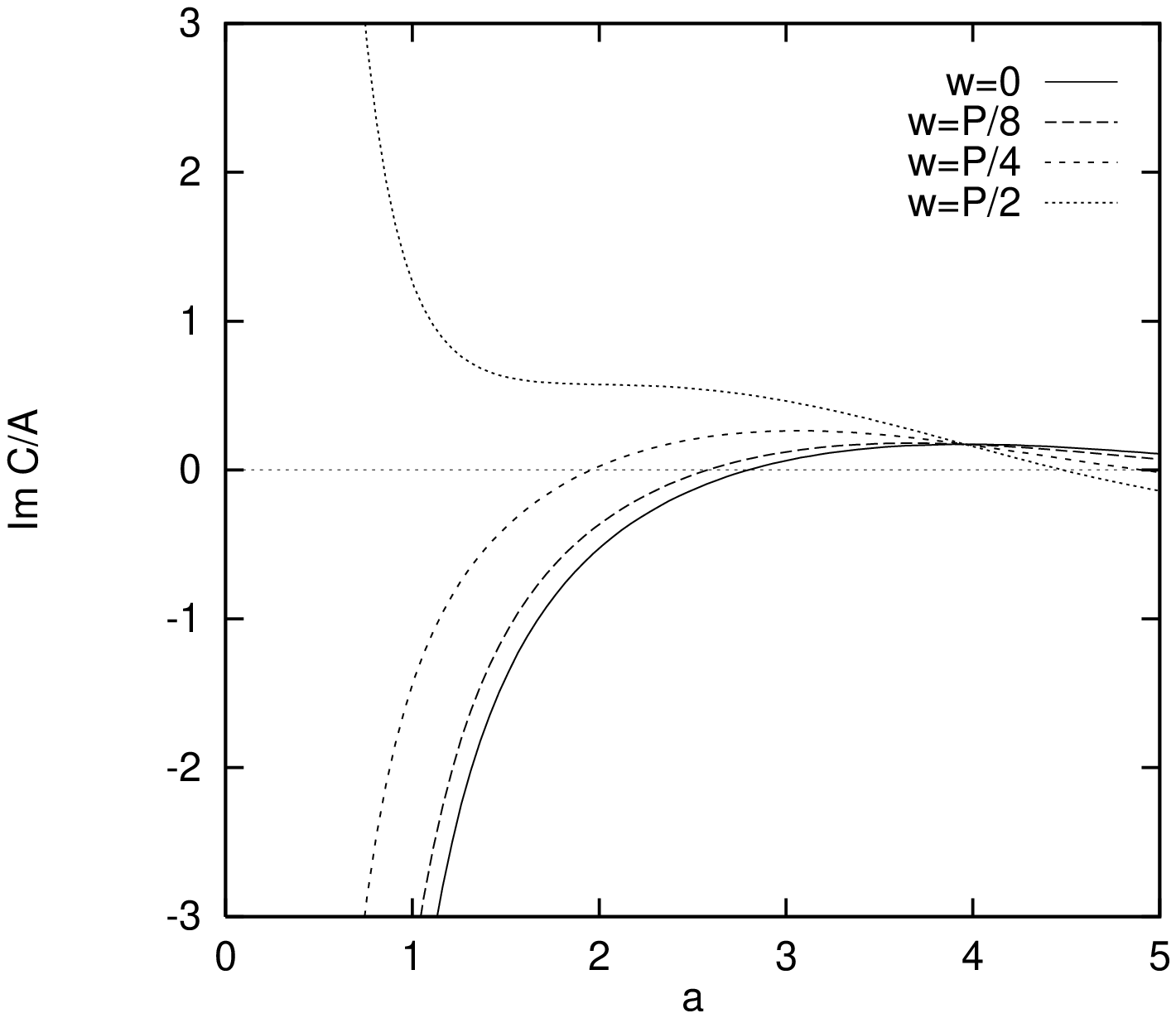, width=8.5cm}
\caption{Dependence of $C$ on $r$}
\label{bild12b}
\end{figure}

The steady-state density matrix 
$\rho^{\rm s s}$ can  be found  from
the Bloch equations \cite{aga} and is known in the literature,
see e.g. Ref. 
\cite{FiTaKi83}. One can also directly employ the Bloch equations in 
Eq. (\ref{16d}) of the Appendix and put 
$\dot{\rho} = \dot{\rho}^{{\rm s s}} = 0$. 
In the Dicke  basis one obtains  for the diagonal elements
\begin{eqnarray} \label{3134}
& & \rho^{\rm s s}_{gg} = {\left(A^2 + \Omega^2 \right)^2 
+ A^2 \, {\rm Re} \, C \left(2 A + {\rm Re} \, C \right) 
+ A^2 \, ({\rm Im} \, C)^2 \over N} \nonumber\\
&& \rho^{\rm s s}_{ss} = {\Omega^2 \left( 2 A^2 + \Omega^2 \right) \over N} ~,
~~~ \rho^{\rm s s}_{aa} = \rho^{\rm s s}_{ee} = {\Omega^4 \over N}
\end{eqnarray}
with the normalization factor
\begin{eqnarray} \label{3135}
N &=& \left(A^2 + 2 \Omega^2 \right)^2 
+ A^2 \, {\rm Re} \, C \left(2 A + {\rm Re} \, C \right) 
+ A^2 \, ({\rm Im} \, C)^2 ~.
\end{eqnarray}

We also need the diagonal elements of normalized reset matrix, 
the density matrix right after an emission. Due to an emission 
the populations of the excited states in the Dicke basis
move down one step to lower levels in proportion to their decay
constants and the previous
ground-state population disappears since it does not contribute to an
emission. Normalization is then achieved by dividing by the trace, tr(.).
This gives 
\begin{eqnarray}\label{30c}
\langle g |\hat {\cal R}(\rho^{\rm s s}) | g \rangle & = & 
\left\{(A + {\rm Re}\, C) \rho^{\rm s s}_{s s} +
(A - {\rm Re}\, C)\rho^{\rm s s}_{a a}\right\}\,/\,{\rm tr(.)}\nonumber \\
\langle s | \hat{\cal R}(\rho^{\rm s s}) | s \rangle & = & 
(A + {\rm Re}\, C) \rho^{\rm s s}_{e e}\,/\,{\rm tr(.)} \nonumber \\
\langle a | \hat{\cal R}(\rho^{\rm s s}) | a \rangle & = & 
(A - {\rm Re}\, C) \rho^{\rm s s}_{e e}\,/\,{\rm tr(.)}
\end{eqnarray}
and $\langle e | \hat{\cal R}(\rho^{\rm s s}) | e \rangle =  0$. 
The complete reset matrix is given in the Appendix.

\begin{figure}[h]
\begin{center}
\psfrag{e}{$\!\!\!\!\!|e\rangle$} 
\psfrag{s}{$\!\!\!\!\!|s\rangle$}
\psfrag{a}{$|a\rangle$}
\psfrag{g}{$|g\rangle$}
\epsfig{file=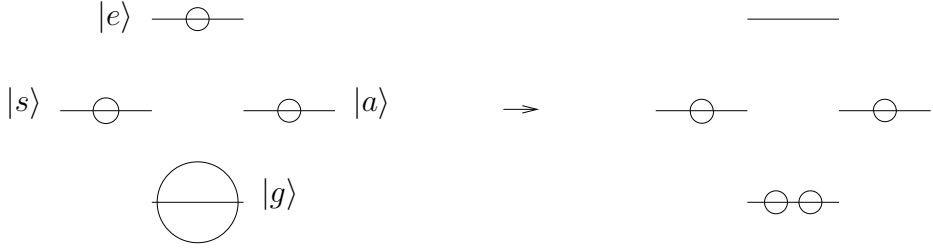,width=12cm}
\caption{Relative populations before and after an 
  emission.}
\label{bes}
\end{center}
\end{figure}

One can immediately draw the following conclusions from these
expressions. \\
(i) For small atomic distance, $k_0r < 2$,
${\rm Im}\, C$ and $N$ become very large. Hence, both for
weak and stronger driving, the 
steady-state population $\rho^{\rm s s}_{gg}$ of the ground
state becomes much larger than that of the excited states, 
and thus the steady-state emission probability is small
in this case. \\
(ii) For strong driving, 
$\Omega \sim A$, the  ratios of steady-state populations of the three
excited states  are of equal order of
magnitude for all atomic distances (since Re$\, C$ does not vary much
and $N$ drops out).
\\
(iii) Right after an emission, the (large) steady-state 
ground-state population is
discarded, the populations of $| s \rangle$ and $| a \rangle$ are
transferred to $| g \rangle$ in the reset matrix and that of $|
e\rangle$ to $| s \rangle$ and $| a \rangle$, all in proportion to
their  appropriate decay
constants. Hence, after an emission and  for $\Omega \sim A$, the
ground-state population and that of the first excited states
have become of similar magnitude (see Fig. \ref{bes} for a qualitative
description).
\\
(iv) After an emission therefore, for small atomic distance and for 
$\Omega \sim A$,
the population of the two first excited states has increased in
relation to the ground-state population. Therefore the probability
density for the next photon right after an emission is higher than
the steady-state emission rate. This means bunching.

This argument for bunching can be extended to  weak driving and small 
distances as follows. From Eq. (\ref{3134}) one sees that in the
steady state only the
numerator of the  ground-state population contains Im$\,C$. Since the
latter increases rapidly for decreasing $r$, as seen from
Fig. \ref{bild12b}, the ratios of the populations of  the excited states with
that of the  ground state  approach 0, while the ratios among the
excited states do not change. After a photon emission
 the upper populations move downwards and the previous large
 ground-state population is discarded. Hence again, after a photon
 emission the ratios of the populations of excited states and ground
 state have increased compared to the steady state 
if the atomic distance is sufficiently small, 
and this means a higher emission probability density, i.e. bunching.

These observations will now be made quantitative. Since $I_{\rm s s}$
is obtained from the level population multiplied by their decay
constants $A \pm {\rm Re}\, C$, one has 
\begin{eqnarray}\label{3136}
I_{\rm ss} & = & (A + {\rm Re}\, C) \rho^{{\rm s s}}_{s s} + (A - {\rm
    Re}\,C) \rho^{{\rm s
    s }}_{a a} + 2 A \rho^{{\rm s s}}_{e e}~.
%\nonumber \\
% & = &{2 \, A \, \Omega^2 \left( 2 \Omega^2 +A \, (A+{\rm Re} \, C) \right)
%\over \left(A^2 + 2\Omega^2 \right)^2 
%+ A^2 \, {\rm Re} \, C \left(2 A + {\rm Re} \, C \right) 
%+ A^2 \,( {\rm Im} \, C)^2} ~.
\end{eqnarray}
Hence the normalization constant tr(.) in Eq. (\ref{30c}) is $I_{\rm
  ss}$. For small atomic distance
$I_{\rm s s}$ becomes very small, due to the small population of the
  excited states. This can be attributed to the detuning
due to the level shift $\hbar\, {\rm Im}\, C$ (see Fig. \ref{bild14}).

For $g(0)$ in Eq. (\ref{34d}) one needs $I(0 ; \hat{{\cal
    R}}(\rho^{\rm s s}))$, the probability density for a new emission
    right after an emission. This is obtained in a similar way as
    $I_{\rm ss}$, 
\begin{eqnarray}\label{52a}
 I(0 ; \hat{{\cal R}}(\rho^{\rm s s})) & = & (A + {\rm Re}\, C) \langle s |
\hat{{\cal R}} (\rho^{\rm s s}) | s \rangle 
 + (A - {\rm Re}\, C) \langle a | \hat{{\cal R}}(\rho^{\rm s s})
| a \rangle + 2 A \langle e | \hat{{\cal R}} (\rho^{\rm s s}) | e
\rangle\nonumber \\
& = & 2\left\{A^2 + ({\rm Re} \, C)^2\right\}\rho^{\rm ss}_{ee}
/I_{\rm ss}
%\frac{\Omega^2 [A^2 + ({\rm Re}\, C)^2]}{A[2
%  \Omega^2 + A^2 + A \,{\rm Re}\, C]}~.
\end{eqnarray}
by Eqs. (\ref{30c}) and (\ref{3136}).  One could also have used 
Eq. (\ref{16z}) of the Appendix. From the behavior of
Re$\,C$ it follows that 
$I(0 ; \hat{{\cal R}}(\rho^{\rm s s}))$ is  of the same order of
magnitude for all atomic distances. This fact is
immediately understood by the observations (ii) and (iii) above.
From Eqs. (\ref{34d}) and (\ref{3134}) - (\ref{52a}) one finally obtains
\begin{eqnarray} \label{3141} 
g(0) &=& {I(0; \hat{{\cal R}}(\rho^{\rm ss}) ) \over I_{\rm ss}} 
\nonumber \\
%&=& {A^2 + ({\rm Re} \, C)^2 \over 2 A^2} \cdot
%{ \left( A^2+2 \Omega^2 \right)^2 +A^2\, {\rm Re} \, C(2A+{\rm Re} \,C)
%+A^2 \, ({\rm Im} \,C)^2 \over \left( 2\Omega^2 
%+ A\, (A+ {\rm Re}\,C) \right)^2} \nonumber \\
&=& {A^2 + ({\rm Re} \, C)^2 \over 2 A^2} \, 
\left[ \, 1+ {A^2 ({\rm Im} \,C)^2-4 \Omega^2 A {\rm Re} \,C 
 \over \left( 2\Omega^2 + A^2+ A {\rm Re}\,C \right)^2} \right] ~.
\end{eqnarray}
 Since $I_{\rm ss}$ becomes small for small $r$  while 
$I(0;\hat{{\cal R}}(\rho^{\rm ss})$
  does not change much with $r$ one has $g(0) > 1$ for small atomic
  distances.
In the last expression  the first factor approaches 1 for small atomic distance
since Re$\, C$ goes to $A$, while the second factor
grows with Im$\, C$. 
\begin{figure}[h]
\psfrag{D=0}{\hspace*{-0.3cm} {\tiny $\vartheta=0$}}
\psfrag{D=p/4}{\hspace*{-0.4cm} {\tiny $\vartheta=\pi/4$}}
\psfrag{D=p/2}{\hspace*{-0.4cm} {\tiny $\vartheta=\pi/2$}}
\psfrag{a        \(a\)}{{\tiny $k_0r$ \hspace*{1cm} (a)}}
\psfrag{a        \(b\)}{{\tiny $k_0r$ \hspace*{1cm} (b)}}
\psfrag{g\(0\)}{{\tiny $g(0)$}}
\hbox to \hsize {
\hss \epsfxsize=8cm \epsfbox {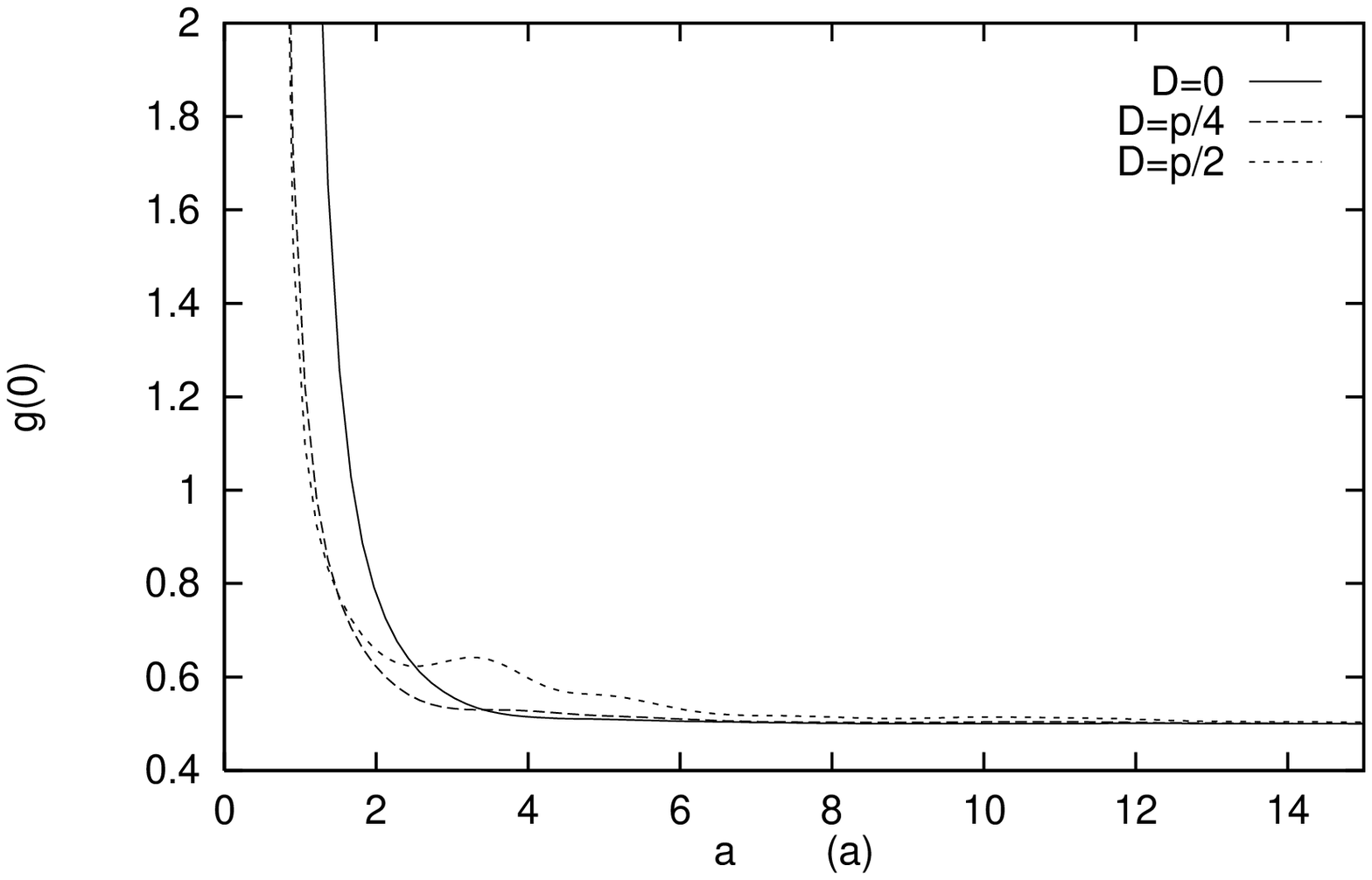} 
\hss \epsfxsize=8cm \epsfbox {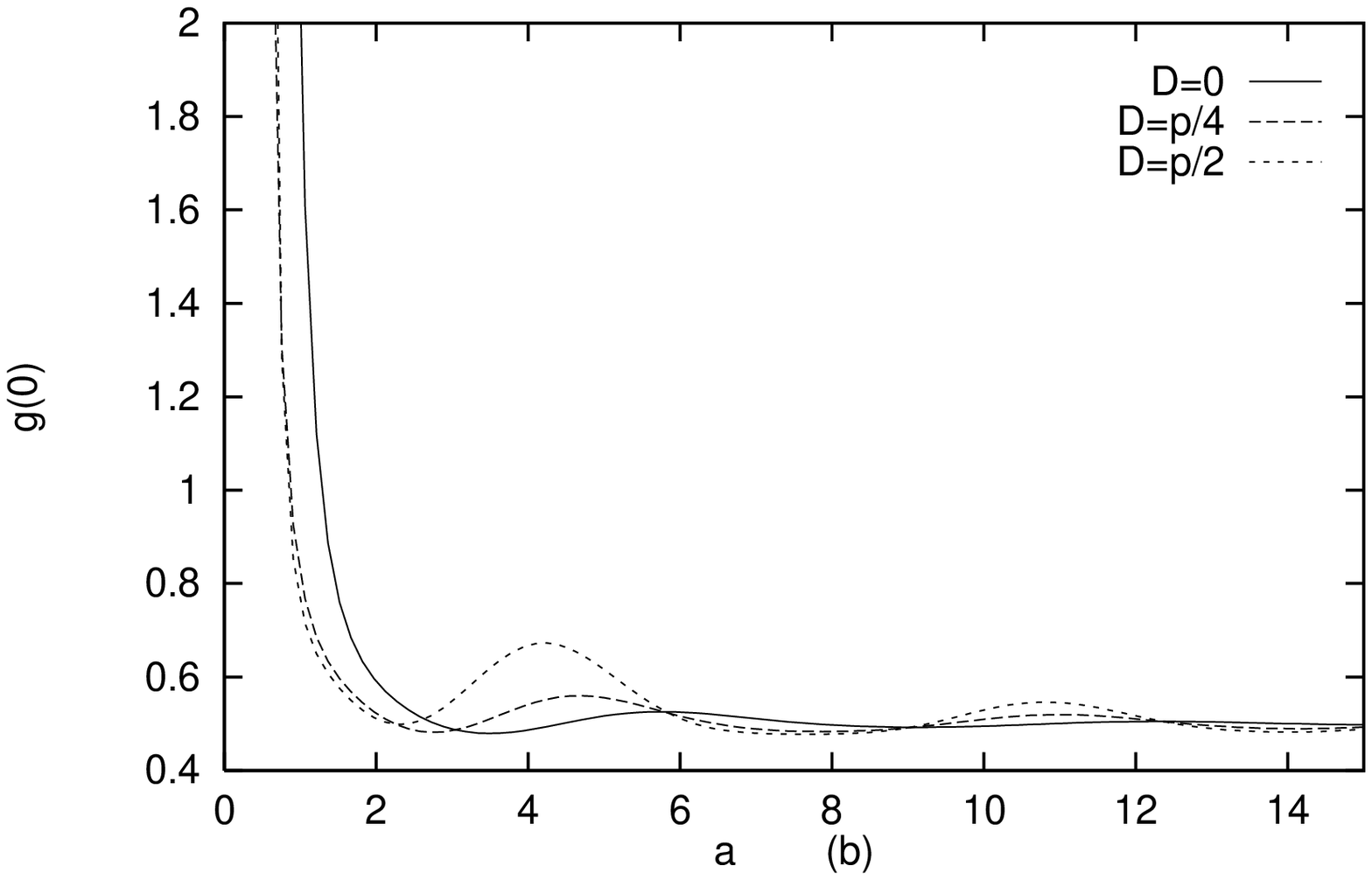} \hss}
\caption{ Photon correlation  $g(0)$ as a function of
$r$. (a) $\Omega=0.1 \, A\,$, (b) $\Omega=0.9 \,
A$}
\label{bild29}
\end{figure}
In particular, for weak driving the terms
involving $\Omega$ can be neglected and one can read off
Fig. \ref{bild12b} that one has bunching below an atomic distance of
about a quarter of the optical wavelength. For strong driving, $\Omega
\sim A$, bunching sets in when the atoms are slightly closer.
 For large atomic distance
$g(0)$ approaches $1/2$ since $C$ approaches $0$. This recovers the
result for two independent atoms. This is plotted in Fig. \ref{bild29}.
%Kap. 4
\section{Discussion}
%Kap. 4
We have investigated bunching and antibunching in the resonance
fluorescence of two atoms as a function of their distance and with
their dipole-dipole interaction taken into account.  
Each atom was treated as a two-level
system and the position of the atoms was kept fixed. The two-atom
system was irradiated by a laser tuned to the transition frequency of
the individual atoms. Retardation effects have been included.

For a single two-level atom antibunching in the resonance
fluorescence is well-known and well understood \cite{CarWa}. After
emission of a photon the atom  is in its ground state and 
 has to be pumped to the excited state before
emitting a new photon. Hence the
probability density for finding a second photon right after an
emission is zero.

For two independent, noninteracting two level atoms one of the atoms
is in its ground state right after an emission while the other is
unchanged. Therefore the probability density for a second photon right
after an emission is half of that in the steady state. This
means $g(0) = 1/2$.

For two interacting two-level atoms the emission statistics depends on
the distance. For atoms far apart the interaction is negligible and one has
antibunching as for two independent atoms. For small atomic
distances one has bunching. The main purpose of this paper was to get a better
understanding of this phenomenon.

Let $| g \rangle \equiv | 1 \rangle | 1 \rangle$ and $| e \rangle
\equiv | 2 \rangle | 2 \rangle$ denote the states where both atoms are
in the ground and excited state, respectively, and let $| s \rangle$ and
$| a \rangle$ be the symmetric and antisymmetric combinations of $| 1
\rangle | 2 \rangle$ and $| 2 \rangle | 1 \rangle$ (Dicke states). 
We first discuss
the case of strong driving. Then,  for 
small distances,  the steady-state ground-state
population is much larger than those in $| s \rangle, | a \rangle$ and
$| e \rangle$ while the populations of the latter 
are of similar (though small) magnitude, as indicated in
Fig. \ref{bes}. The reason
for the small population of the latter is easily understood through the
level shift of $| s \rangle$ and $| a \rangle$ due to the dipole force
(see Fig. \ref{bild14}). 
The reason for the similar magnitude of the population of 
$| s \rangle$ and $| a
\rangle$ with $| e \rangle$ has been attributed to 
two-photon processes connecting $| g \rangle$ with $| e \rangle$
\cite{VaAga}. Now, once a system has emitted a photon the population of
$|e \rangle$  is transferred to $| s \rangle$ and $| a
\rangle$ and the population of the two latter to $| g \rangle$, in
proportion to the respective decay constants  and with
ensuing normalization. The previous population of $| g
\rangle$ has disappeared since it does not contribute to the
emission. Thus right after an emission the populations of $| s \rangle, | a
\rangle$ and $| g \rangle$ are suddenly of similar magnitude while
before the emission the population of the ground state was much
larger. Hence right after an emission the probability density for
finding another 
photon has increased when compared to that preceding the emission,
i.e. compared to the steady state emission rate. This means bunching
for small distances and strong driving. 

For weak driving the mechanism is in principle the same. Although
in the steady state the populations of the excited states now are no longer of 
similar magnitude the ground-state population   increases  with 
decreasing distance much faster than the population difference between
the excited states. This means that the population of $|e \rangle$ is
not too small when compared to the population of  $|s \rangle$
and $|a \rangle$. Therefore after an emission, when the populations
have moved down one step, the combined population of $|s \rangle$ and
$|a \rangle $ has increased. This means a higher
emission probability density than in the steady state, i.e. bunching.

We have shown that when decreasing the atomic distance 
the transition from antibunching to 
bunching sets in at a distance of
about a quarter of the optical wavelength, for weak driving slightly
sooner than for strong driving.

It is instructive to see why the same argument gives antibunching for two
independent, non-interacting atoms. First, for strong driving, the two
levels of each individual atom are populated by approximately $1/2$ so that the
population of $| g \rangle, | s \rangle, | a \rangle$, and $| e
\rangle$ are $1/4$ each. Then, after an emission, the ratios of the
populations of $| g \rangle, | s \rangle$,  $| a \rangle$ and $| e \rangle$ are
$\frac{1}{2}~:~\frac{1}{4}~:~\frac{1}{4}~:~0$, as inherited from those of
$| s \rangle, | a \rangle, | e \rangle$, prior to the emission and in
proportion to the decay rates. Thus
the probability density for  a next emission 
is only one half of that in the steady state. 
On the other hand, for weak driving the
ground-state population is much larger than that of the excited
states. Is this situation not similar to that of interacting atoms?
Not quite, since although the populations of $| s \rangle$ and $| a
\rangle$ are small and of similar magnitude, that of $| e \rangle$ is
of the order of the product of the latter and therefore an order of
magnitude smaller. Thus after an emission the ground state population
is still much larger than that of the excited states, and there is no
increase in the emission probability.

The analysis can be carried over to the more general case where the
dipole moments are not parallel and where the laser is detuned and its
phase is different for the two atoms. The necessary tools are given in
the Appendix. Also the case of degenerate
upper level can be treated. The results  \cite{Be} 
are similar to those obtained
above.

To conclude, we have traced the appearance of bunching in the
resonance fluorescence of a  driven system of two two-level
atoms with dipole-dipole interaction and at small distances 
to two causes, one the level
populations of the steady-state density matrix, the other the change
in the state right after the emission of a photon. A similar analysis
can in principle also be applied to other systems, e.g. to a single
atom in a three-level cascade configuration.
%App.
\section*{Appendix}
We consider two atoms fixed at positions ${\bf
  r}_i$ and each with two levels, $|1\rangle_i$ and $|2\rangle_i, i =
  1,2$, with energy difference $\hbar \omega_0$. We define
  operators $S^\pm_i$ in the two-atom Hilbert space by
$S^+_i = |2\rangle_{ii}\langle 1|$ and $S_i^-
= |1 \rangle_{ii} \langle 2|$.
The dipole moment of the $i$-th atom is
$ {\bf D}^{(i)}_{12} = ~_i\langle 1|{\bf X}|2\rangle_i$.
For the laser we take zero detuning and $
{\bf E}_L ({\bf r},t) = {\rm Re}\left\{ {\bf E}_0~ e^{{\rm i}({\bf
      k}_L \cdot {\bf r} - \omega_0 t)} \right\}$.
Making the usual rotating-wave approximation and going over to the
interaction picture the interaction Hamiltonian becomes
\begin{eqnarray} \label{26}
H_{\rm I} = \sum_{i=1}^2 \sum_{{\bf k},s} \hbar \left[
g_{{\bf k},s}^{(i)} a_{{\bf k},s} \,  
{\rm e}^{{\rm i} (\omega_0-\omega_k) t} \, 
{\rm e}^{{\rm i} {\bf k} \cdot {\bf r}_i} S_i^+ 
+ \, {\rm h.c.} \, \right] + H_L ~,
\end{eqnarray}
with the coupling constants 
\begin{eqnarray}  \label{25}
g_{{\bf k},s}^{(i)} &=& {\rm i} e 
\left( {\omega_k \over 2 \epsilon_0 \hbar L^3} \right)^{1/2} 
\left( {\bf D}^{(i)}_{12},
{\bf \epsilon}_{{\bf k},s} \right),
\end{eqnarray}
laser part 
$
H_L = ~\frac{\hbar}{2}~\sum^2_{i = 1} \left\{ \Omega_i S_i^+ +
 \Omega_i^\star S_i^- \right\}
$
and Rabi frequencies
$
\Omega_i =~\frac{e}{\hbar}~{\bf D}_{1 2}^{(i)} \cdot {\bf E}_0~e^{i
  {\bf k}_L \cdot {\bf r}_i}$.
The operator $H_I$ contains the dipole-dipole interaction of the two
atoms as seen from the Bloch equations or from the conditional
Hamiltonian between emissions, as explained further 
below. In the above Power-Zienau formulation this interaction is
due to photon exchange \cite{aga}.

{\em Reset matrix}.
The reset operation gives the state or
density matrix right after a photon detection. In a basis in which the atomic
damping is diagonal, as for the Dicke states, the diagonal states
immediately can be written down, as in Eq. (\ref{30c}).
For a general $N$-level system
the reset matrix has been derived in Refs. \cite{He,HeSo}. For a system
consisting of two or more atoms the derivation has to slightly
modified since in this case the  field operator ${\bf E}$ appear with
different position arguments.

Let at time $t$ the state of the combined system, atoms plus quantized
radiation field, be given by
$
|0_{\rm ph} \rangle\, \rho\,\langle 0_{\rm ph}|$,
i.e. the atomic system is described by the density matrix $\rho$ and
there are no photons (recall that the laser field is treated
classically). If at time $t + \Delta t$ a photon is detected (but not
absorbed) the combined system is in the state
\begin{equation}\label{14a}
I\!\!P_{\!>} U_I(t + \Delta t, t) |0_{\rm ph} \rangle \,\rho \,\langle
0_{\rm ph}|
U_I^\dagger (t + \Delta t, t) I\!\!P_{\!>}
\end{equation}
where
$I\!\!P_{\!>} = {\bf 1} - |0_{\rm ph} \rangle {\bf 1}_A \langle 0_{\rm
  ph}|$
is the projector onto the one or more photon space (since $\Delta t$
is small one could directly take the projector onto the one-photon
space). The probability for this event is the trace over Eq.
(\ref{14a}). For the state of the atomic system it is irrelevant
whether the detected photon is absorbed or not (intuitively the photon
travels away and does no longer interact with the atomic
system). Hence after a photon detection at time $t + \Delta t$ the
non-normalized 
state of the atomic system alone, denoted by ${\cal R} (\rho) \Delta
t$, is given by a partial trace over the photon space, 
\begin{equation}\label{215}
{\cal R} (\rho) \Delta t = {\rm tr_{ph}} \left( I\!\!P_{\!>} U_I (t
+ \Delta t,t) |0_{ph} \rangle \,\rho \,\langle0_{ph}|U^\dagger_I (t +
\Delta t, t)
 I\!\!P_{\!>} \right)~.
\end{equation}
We call ${\cal R} (\rho)$ the non-normalized reset state \cite{He}. 
Proceeding as in Refs. \cite{He,HeSo} and  using
perturbation theory one obtains \cite{Be}
\begin{eqnarray} \label{216}
{\cal R} (\rho) 
&=& {1 \over 2}\left( C_{12}^* +C_{21} \right) S_1^-\rho S_2^+
+{1 \over 2}\left( C_{12} + C_{21}^* \right) S_2^-\rho S_1^+
\nonumber \\
& & + A \left(S_1^- \rho S_1^+ +S_2^- \rho S_2^+ \right) 
\end{eqnarray} 
with the $r$ dependent constants
\begin{eqnarray} \label{210}
C_{ij} &=& {3A \over 2} \, {\rm e}^{{\rm i} k_0r} \Bigg[
{1 \over {\rm i}k_0r} \left( \left(\hat{{\bf D}}_{12}^{(i)},
\hat{{\bf D}}_{12}^{(j)}\right)
- \left(\hat{{\bf D}}_{12}^{(i)}, \hat{{\bf r}}\right)
\left(\hat{{\bf r}}, \hat{{\bf D}}_{12}^{(j)}\right) \right)
\nonumber \\
& & +\left( {1 \over (k_0r)^2} -{1 \over {\rm i}(k_0r)^3} \right) 
\left( \left(\hat{{\bf D}}_{12}^{(i)},
\hat{{\bf D}}_{12}^{(j)}\right)
-3 \left(\hat{{\bf D}}_{12}^{(i)}, \hat{{\bf r}} \right)
\left(\hat{{\bf r}}, \hat{{\bf D}}_{12}^{(j)} \right) \right) 
\Bigg]~. 
\end{eqnarray}
Here $\hat{~}$  denotes vectors normalized to 1, A is the Einstein
coefficient, and
${\bf r} = {\bf r}_2 - {\bf r}_1$.
In the case of equal dipole moments one has $C_{12}=C_{21}\equiv C$
which was depicted in Fig. \ref{bild12b}, with $\vartheta$ defined by
$\cos^2 \vartheta = \left| \left( {\bf D}_{1 2} , {\bf
        r} \right) \right|^2/r^2 D_{12}^2
$.
The normalized reset state is 
$\hat{{\cal R}}(\rho) \equiv {\cal R}(\rho)/{\rm tr} {\cal R}(\rho)$.

By Eq. (\ref{14a}) the normalization of ${\cal R}(\rho)$ is such
that tr$_A {\cal R}(\rho) \Delta t$ is the probability for a photon
detection at time $t + \Delta t$ if the (normalized) state of the
atomic system at time $t$ is $\rho$. Hence one has 
 for
the probability density  $I$ of Section 2
\begin{equation}\label{16z}
I(t;\rho(0)) = {\rm tr}\, {\cal R}(\rho(t))~.
\end{equation}

The laser field does not appear in the reset state, just as in the
case of a single atom \cite{He,HeSo}, since its effect 
during the short time $\Delta t$  is negligible.
By a simple calculation one checks that Eq. (\ref{216}) can be written
as 
\begin{equation}\label{16a}
{\cal R} (\rho) = \left(A + ~\frac{1}{2}~|C_{1 2} + C^\star_{2 1}|\right)
R_+ \rho R^\dagger_+ + 
\left(A - ~\frac{1}{2}~|C_{1 2} + C_{21}^\star |\right) R_- \rho R_-^\dagger
\end{equation}
where $R_\pm = \left(S_1^- \pm e^{i \varphi} S_2^- \right) / \sqrt{2}$
and $\varphi$ is the argument of $C_{1 2} + C_{1 2}^\star$. From
Eq. (\ref{210}) one can check that $A \ge\frac{1}{2} |C_{1 2} + C_{2 1}^\star
|$. If $\rho$ is a pure state, $\rho = | \psi \rangle \langle \psi |$
say, then $R_\pm \rho R_\pm^\dagger$ are also pure states. This
decomposition of ${\cal R}(\rho)$ is advantageous for simulations of
trajectories. 

{\em Conditional Hamiltonian and waiting times}.
In the quantum jump approach \cite{HeWi,Wi,He,PleKni}, the time development
of an atomic system is described by a conditional non-hermitian
Hamiltonian $H_{{\rm cond}}$, which gives the time development between
photon emissions, and by a reset operation which gives the state or
density matrix right after an emission. For a general $N$-level system
these have been derived in Refs. \cite{He,HeSo}. The derivation of the
former is adapted here to a system of two atoms.

As explained in Refs. \cite{HeWi,He,HeSo,PleKni}, $H_{{\rm cond}}$ is
of the general form
$
H_{{\rm cond}} = H_A + H_L + \Gamma
$
where $\Gamma$ is an atomic damping operator. In a basis in which $\Gamma$ is
diagonal the diagonal terms are just the decay constants of the
corresponding states. If these (dressed) states are known 
$H_{{\rm cond}}$ can immediately be
written down. In this way one can obtain $H_{{\rm cond}}$ for parallel
dipole moments in the Dicke basis. In the general case it is 
obtained (in the interaction picture) from the short-time development
under the condition of no emission, i.e. from the relation
\begin{equation*}\label{7a}
{\bf 1} -~\frac{\rm i}{\hbar}~H_{{\rm cond}} \Delta t = \langle 0_{ph}|
U_I(\Delta t, 0) |0_{ph}\rangle
\end{equation*}
where the r.h.s. is evaluated in second order perturbation theory for
$\Delta t$ intermediate between inverse optical frequencies and
atomic decay times. In a similar way as for a single atom
\cite{HeWi,He,HeSo} one obtains for two two-level atoms \cite{Be}
\begin{eqnarray} \label{29}
H_{\rm cond} &=& \frac{\hbar}{2{\rm i}} \left[ 
A\left(S_1^+S_1^- + S_2^+S_2^-\right) +
C_{12} \, S_1^+S_2^- +C_{21} \, S_2^+S_1^-\right] + H_L
\end{eqnarray}
with the $r$ dependent constant $C_{ij}$ given by Eq. (\ref{210}).
Between emissions the time development is given by
$U_{{\rm cond}} (t, 0) = \exp \left\{ - {\rm i} H_{{\rm cond}} t/\hbar \right\}
$
which is non-unitary since $H_{{\rm cond}}$ is non-hermitian. The
corresponding decrease in the norm of a vector is connected to the
waiting time \cite{CohDal} for emission of a (next) photon. If at $t = 0$ the
initial atomic state is $| \psi \rangle$ then the probability $P_0(t)$ to
observe {\em no} photon until time $t$ 
by a broadband counter (over all space) is
given by \cite{HeWi,He,HeSo}
\begin{equation}\label{212}
P_0 (t;  |\psi \rangle) = \|U_{{\rm cond}}(t, 0) |\psi \rangle \|^2~,
\end{equation}
and the probability density $w_1$ of finding the first photon at time
$t$ is 
\begin{equation}\label{213}
w_1 (t; |\psi \rangle) = -~\frac{d}{dt}~P_0 (t; |\psi \rangle)~.
\end{equation}
For an initial density matrix instead of $| \psi \rangle$ the
expressions are analogous, with a trace instead of a norm squared in
Eq. (\ref{212}). For $t=0$ one must have $w_1(0)=I(0)$ since for short
times any photon must be the first. This identity is easily checked by
means of Eqs. (\ref{212}), (\ref{16z}) and (\ref{216}).

For equal dipole moments and without laser the conditional Hamiltonian is
diagonal in the  Dicke basis. 
$A \pm{\rm Re}\, C$ describes the decay rates of
$| s \rangle$ and $| a \rangle$ to $| g \rangle$, while $\pm \hbar
~{\rm Im}\, C$ can be viewed as a level shift. The state $|e \rangle$ can
decay to both $| s \rangle$ and $| a \rangle$, with respective decay
rates $A \pm {\rm Re}\, C$. This also follows from the Bloch equations
and is indicated in Fig. \ref{bild14}. 
From this the well-known fact 
follows that two atoms with dipole interaction can decay
faster or slower than two independent atoms (super- and sub-radiance
\cite{Brewer1}). When $r \rightarrow 0$, ${\rm Re}\, C$ approaches $A$ so
that $| a \rangle$ can no longer decay while $| s \rangle$ decays with
$2 A.$

{\em Trajectories and Bloch equations}.
Starting at $t = 0$ with a pure state, the state develops
according to $U_{{\rm cond}}$ until the first emission at some time
$t_1$, determined from $w_1$ in Eq. (\ref{213}). Then the state is
reset according to Eq. (\ref{216}) to a new density matrix (which has
to be normalized), and so on.

The decomposition of ${\cal R}(\rho)$ in Eq. (\ref{16a}) allows one,
however, to work solely with pure states which is numerically much
more efficient. One can start with a pure  state $| \psi \rangle$,
develop it with $U_{{\rm cond}}$ until $t_1$ to the (non-normalized)
$|\psi(t_1)\rangle$, reset to one of the pure states
$
R_\pm |\psi(t_1)\rangle/\| \cdot \|
$
with relative probabilities given by the factors $A\pm~\frac{1}{2} |
C_{1 2} + C_{2 1}^\star |$ appearing in Eq. (\ref{16a}), and so
on. The waiting time distributions are not changed by this procedure.

Quite generally the ensemble of such trajectories yields the Bloch
equations \cite{He}. With the reset matrix this is easily seen as
follows. If an ensemble of systems of two two-level atoms has a
density matrix $\rho(t)$ at time $t$ then at time $t + \Delta t$ one
has two sub-ensembles, one with a photon emission, the other with
none. The former has relative size tr$\,{\cal R}(\rho (t)) \Delta t$, by the
remark after Eq. (\ref{216}), while the latter is obtained by means of
$U_{{\rm cond}} (t + \Delta t, t) = {\bf 1}_A -~\frac{i}{\hbar}
H_{{\rm cond}} \Delta t$. This immediately gives
\begin{equation}\label{16d}
\dot{\rho} = -~\frac{\rm i}{\hbar}~[H_{{\rm cond}} \rho - \rho
H^\dagger_{{\rm cond}}] + {\cal R}(\rho)~. 
\end{equation}
Inserting from Eqs. (\ref{29}) and (\ref{216}) one obtains the Bloch
equations for two two-level atoms. They agree with those derived by
Agarwal \cite{aga}. 
From this expression it is evident that  $H_{\rm cond}$ or the 
reset matrix can be immediately determined  
if  the Bloch equations and the reset matrix or, respectively, 
$H_{\rm cond}$  are explicitly known.

\end{document}